\pdfoutput=1
\documentclass[11pt,colorinlistoftodos]{article}
\usepackage[utf8]{inputenc}
\usepackage{amssymb}
\usepackage{bm}
\usepackage{booktabs}
\usepackage{array}
\usepackage{latexsym}
\usepackage{graphicx}
\usepackage{color}
\usepackage{verbatim}
\usepackage{enumerate}
\usepackage{chngpage}
\usepackage{mathrsfs}
\usepackage{euscript}
\usepackage{psfrag}
\usepackage{soul}
\usepackage{empheq}
\usepackage{tikz}
\usepackage{amsmath}
\usepackage{amsfonts}
\usepackage{amsthm}
\usepackage{titlesec}
\usepackage{setspace}
\usepackage{tensor}
\usepackage{physics}
\usepackage{mciteplus}
\usepackage{todonotes}  
\usepackage{mathtools}
\usepackage[nosort]{cite}
\titleformat{\paragraph}
{\normalfont\normalsize\bfseries}{\theparagraph}{1em}{}
\titlespacing*{\paragraph}
{0pt}{3.25ex plus 1ex minus .2ex}{1.5ex plus .2ex}

\usepackage[colorlinks=true,
            linkcolor=blue,
            urlcolor=blue,
            filecolor=blue,
            citecolor=blue,
            pdfstartview=FitH,
            pdfpagemode=UseNone,
            bookmarksopen=true]{hyperref}
\usepackage[all]{hypcap}

\definecolor{cardinal}{rgb}{0.6,0,0}
\definecolor{darkgreen}{rgb}{0,0.4,0}
\definecolor{golden}{rgb}{0.92,0.7,0}
\definecolor{midnight}{rgb}{0,0,0.5}
\definecolor{darkblue}{rgb}{0,0,0.7}
\definecolor{purple}{rgb}{0.5,0,0.5}

\def\n_1{{l}}

\newcommand{\Vt}{V_{11}(T_2)}
\numberwithin{equation}{section}

\begin{document}

\phantom{AAA}
\vspace{-10mm}

\begin{flushright}
\end{flushright}

\vspace{1.9cm}

\begin{center}
{\huge \bf Positive spin-induced Quadrupole Moments}

\medskip

{\huge \bf of Bobev-Ruef Solitons}

\vspace{1cm}

{\large\bf Ang\`ele Lochet}

\vspace{1cm}

Institut de Physique Th\'eorique,\\
Universit\'e Paris-Saclay, CEA, CNRS,\\
Orme des Merisiers, Gif-sur-Yvette, 91191 CEDEX, France\\[12pt]

{\footnotesize\ttfamily angele.lochet@ipht.fr}

\vspace{2.2cm}

\textsc{Abstract}
\end{center}

\begin{adjustwidth}{3mm}{3mm}
\vspace{-1.2mm}
\noindent
We calculate the spin-induced quadrupole moment of Bobev-Ruef solitons: smooth horizonless solutions of eleven-dimensional supergravity obtained by adding fluxes on a Euclidean Einstein-Maxwell base space that generalizes the Euclidean-Kerr-Taub-Bolt space. We identify the solitons that descend to rotating four-dimensional solutions with flat-space asymptotics and no closed timelike curves, and we compute their mass, angular momentum, quadrupole moment and Page charges. The quadrupole moment takes the simple form $M_2=M(\alpha^2-b^2)$, where $\alpha$ is the rotation parameter of the base and $b$ the displacement of the center of mass, and it is positive in a finite range of fluxes, which contains the solutions that do not run. The multipole ratio $M_2M/J^2$ can reach values around two, much larger than for the Running-Kerr-Taub-Bolt solutions, and, unlike for these solutions, a positive quadrupole moment can arise from base spaces whose signature never changes. We also find a mass formula, check that the BPS bound is respected, and show that the $E_{7(7)}$ quartic invariant of the charges is $-16\,J^2$. Finally, we show that these solitons have a mass below the extremal black hole mass with the same charges and with fixed angular momentum: no black hole has their charges and mass.
\end{adjustwidth}

\thispagestyle{empty}
\newpage

\tableofcontents

\section{Introduction}
\label{sec:intro}

Gravitational-wave observations of binary black-hole mergers by LIGO, Virgo and KAGRA give access not only to the masses and spins of the merging objects, but also to their spin-induced quadrupole moment, $M_2$ \cite{Krishnendu:2017shb}. For a Kerr black hole $M_2=-J^2/M$: the black hole becomes oblate when it spins, bulging at its equator, as do objects made of ``normal'' matter, such as neutron stars. Writing $M_2=-\kappa\,J^2/M$, with $\kappa=1$ for Kerr, an object with $\kappa<0$, that is with a positive quadrupole moment, would instead become prolate when spinning: it would elongate along the rotation axis. Such an object would be a clear sign of new physics.

For most events the quadrupole moment is only weakly constrained, and the data of the third observing run were compatible with, and even appeared to favor, a positive quadrupole moment \cite{LIGOScientific:2021sio}. This has changed with the event GW241011, whose primary black hole spins rapidly and is much heavier than its companion. The spin-induced quadrupole moment of this black hole differs from the Kerr value by $\delta\kappa_1=0.10^{+0.82}_{-0.82}$, which is the most stringent constraint on a spin-induced quadrupole moment to date \cite{LIGOScientific:2025brd}. This object is therefore oblate, like a Kerr black hole, and this measurement already excludes, for example, rotating boson stars as a model of it \cite{Krishnendu:2025rud}. However, it constrains only one compact object, and the question of which objects can have a positive quadrupole moment, and why, remains open.

In String Theory, one can address this question using black-hole microstate geometries: smooth horizonless solutions of supergravity that have the same charges, mass and angular momentum as black holes \cite{Mathur:2005zp,Bena:2007kg,Bena:2022rna,Bena:2022ldq}. These solutions are supported against collapse by their nontrivial topology and fluxes \cite{Gibbons:2013tqa}, and from a four-dimensional perspective they contain very unusual matter. For example, some centers of the smooth supersymmetric solutions correspond, when reduced to four dimensions, to D6 branes of negative mass and negative charge \cite{Bena:2005va,Berglund:2005vb,Bena:2007kg}, and thus can be thought of as ``lighter'' than the vacuum.

The largest families of microstate geometries correspond to supersymmetric black holes \cite{Bena:2006kb,Bena:2007qc,Bena:2010gg,Bianchi:2016bgx,Bianchi:2017bxl,Heidmann:2017cxt,Bena:2017fvm,Avila:2017pwi,Tyukov:2018ypq,Warner:2019jll,Rawash:2022sum} and to extremal non-supersymmetric black holes \cite{Goldstein:2008fq,Bena:2009ev,Bena:2009en,DallAgata:2010srl,Vasilakis:2011ki,Heidmann:2018mtx}, which have infinite AdS$_2$ throats. More recently, smooth horizonless solutions with the charges of non-extremal black holes have been constructed systematically \cite{Bah:2020ogh,Bah:2020pdz,Bah:2021owp,Bah:2021rki,Bah:2022pdn,Heidmann:2021cms,Heidmann:2022zyd,Bah:2022yji,Bah:2023ows,Heidmann:2023kry,Chakraborty:2025ger,Dima:2025tjz}, but all of them are static, and one cannot talk about their spin-induced quadrupole moment.

Adding rotation to smooth horizonless solutions is notoriously difficult. The first rotating examples, the JMaRT solutions \cite{Jejjala:2005yu}, their multi-center generalizations \cite{Bossard:2014yta,Bossard:2014ola,Bena:2015drs,Bena:2016dbw,Bossard:2017vii} and their four-dimensional analogue \cite{Giusto:2007tt}, have an angular momentum above the black-hole bound. Rotating solutions with four-dimensional asymptotics can be obtained by adding self-dual fluxes on a Euclidean Kerr-Taub-Bolt space: these are the Running-Bolt solutions \cite{Bena:2009qv,bena2025positivespininducedquadrupolemoment}. Rotating topological stars were then constructed, first with non-standard asymptotics \cite{Bianchi:2025uis}, and then with standard Kaluza-Klein asymptotics \cite{HPS}. There has also been recent progress in the systematic construction of rotating solutions: an algebraic method now allows one to add rotation to any static solution of five-dimensional minimal supergravity \cite{Chakraborty:2026aah}, and offers a new tool to build rotating smooth horizonless geometries. However the general major issue with these rotating mooth horizonless solutions, is that they tend to lie outside the black-hole regime.

The multipole moments of smooth horizonless solutions have been computed mostly for supersymmetric and almost-BPS solutions \cite{Bena:2020see,Bianchi:2020bxa,Bena:2020uup,Bianchi:2020miz,Ganchev:2022vrv,Bah:2021jno}, where they can be both positive and negative. Among rotating non-extremal solutions, two families are known to have a positive spin-induced quadrupole moment. In a previous paper \cite{bena2025positivespininducedquadrupolemoment}, we identified the Running-Kerr-Taub-Bolt solutions that descend to four-dimensional rotating solutions with flat asymptotics, and found that their spin-induced quadrupole moment is positive when the bolt does not run or runs slowly. The asymptotically-flat rotating topological stars of \cite{HPS} can also have a positive quadrupole moment, in a small region of their parameter space. In both cases, the four-dimensional base space of the solution has regions where its signature changes, and this was proposed as the origin of the positive quadrupole moment: these regions signal ingredients of negative mass, which are lighter than the vacuum, and spinning them makes the object thinner rather than flatter \cite{bena2025positivespininducedquadrupolemoment,HPS}. 

It is natural to ask whether this unusual behavior is specific to these solutions, or whether it is a more generic feature of smooth horizonless solutions of String Theory. In this paper we answer this question for a larger family of solutions: the Bobev-Ruef solitons \cite{Bobev:2009kn}. These solutions are built using the floating-brane construction \cite{Bena:2009fi}, in which the four-dimensional Euclidean base space is not Ricci-flat, but solves the Euclidean Einstein-Maxwell equations, and the Maxwell field of the base becomes one of the magnetic fluxes of the eleven-dimensional solution. The base space of Bobev and Ruef generalizes the Euclidean-Kerr-Taub-Bolt space by an electric and a magnetic parameter, $q$ and $p$, and the Bobev-Ruef solitons reduce to the Running-Kerr-Taub-Bolt solutions when $p=q$. Determining the quadrupole moment of these solitons was one of the open questions of \cite{bena2025positivespininducedquadrupolemoment}.

As for the Running-Kerr-Taub-Bolt solutions, not every smooth Bobev-Ruef soliton descends to a four-dimensional solution with flat asymptotics. The absence of conical singularities and of Dirac strings imposes a relation between the periodicities of the Kaluza-Klein circle and of the angle along which the solution rotates, and this relation must be compatible with the $2\pi$-periodicity of this angle. The compatibility condition is labeled by an integer, $\ell$. The main difference with the Running-Kerr-Taub-Bolt case is that the base space has one more parameter, $\delta\equiv q^2-p^2$. Hence, for a given $\ell$, the regular base spaces are no longer isolated points but form lines in the space of parameters.

We compute the mass, $M$, the angular momentum, $J$, and the spin-induced quadrupole moment, $M_2$, of the Bobev-Ruef solitons, and find that the quadrupole moment takes the very simple form 
\begin{equation}
	M_2=M\,(\alpha^2-b^2)\, , 
\end{equation}
where $\alpha$ is the rotation parameter of the base and $b$ is the shift needed to put the center of mass at the origin. This relation also holds for the Running-Kerr-Taub-Bolt solutions. For all the regular solutions we studied, $M_2$ is positive for a finite range of fluxes, which always contains the fluxes for which the solution does not run, and it becomes negative and diverges as the running speed approaches the speed of light. The dimensionless multipole ratio, $R\equiv M_2M/J^2$, can be much larger than for the Running-Kerr-Taub-Bolt solutions: in one of our examples it reaches $R\simeq 2.07$, while it stays below $0.33$ for other computed solutions with $\ell=1,2,3$ of \cite{bena2025positivespininducedquadrupolemoment} (for comparison, the Kerr black hole has $R=-1$).

Moreover, unlike the Running-Kerr-Taub-Bolt base spaces, some of the Bobev-Ruef base spaces with flat four-dimensional asymptotics have a positive-definite metric everywhere. Hence, a positive spin-induced quadrupole moment does not require the base space to be ambipolar, and the explanation in terms of ingredients of negative mass must be refined.

Finally, we compute the Page charges of the solutions. We find a mass formula that generalizes the one of the Running-Kerr-Taub-Bolt solutions \cite{BenaLochetPage}, check that the BPS bound is satisfied, and find that the quartic invariant of the charges $E_{7(7)}$ is related to the four-dimensional angular momentum by $E_{7(7)}=J_{4}=-16\,(G_4J)^2$, for all values of the fluxes. Using these charges, we show that for fixed charges and angular momentum, the solitons we studied have a smaller mass than the extremal black hole mass with the same charges and angular momentum, so that no black hole has their charges and mass. This is consistent with the fact that the rotating smooth horizonless solutions known so far lie outside the black-hole regime \cite{Chakraborty:2026aah}.

In Section~\ref{sec:base} we review the Bobev-Ruef base space, the conditions for having flat four-dimensional asymptotics, and the fluxes and warp factors of the solution. In Section~\ref{sec:multipoles} we compute the mass, angular momentum and quadrupole moment. In Section~\ref{sec:regular} we identify the solutions free of closed timelike curves and discuss the sign of their quadrupole moment. In Section~\ref{sec:page} we compute the Page charges, discuss the relations between them, and show that there is no black hole with the same charges and mass. We conclude in Section~\ref{sec:conclusion}.

\section{The base metric}
\label{sec:base}
The Bobev-Ruef solitons \cite{Bobev:2009kn} are floating-brane solutions \cite{Bena:2009fi}: instead of a Ricci-flat base space, one starts from a four-dimensional Euclidean space that solves the Einstein-Maxwell equations, and the Maxwell field of this base becomes one of the three magnetic fluxes of the eleven-dimensional solution. The base space used by Bobev and Ruef has, besides the mass parameter $m$, the NUT charge $N$ and the rotation parameter $\alpha$, an electric and a magnetic parameter, $q$ and $p$.
The four-dimensional base metric is:
\begin{equation}
\label{eq:BRbase}
    ds_4^2 = \dfrac{\Sigma}{\Delta} dr^2 + \Sigma d\theta^2 + \dfrac{\sin^2\theta}{\Sigma}\left(\alpha d\tau+ P_r d\phi\right)^2 + \dfrac{\Delta}{\Sigma}\left(d\tau + P_\theta d\phi\right)^2
\end{equation}

with:

\begin{align*}
P_r &= r^2 - \alpha^2 - \frac{N^4}{N^2 - \alpha^2}, & P_\theta &= 2N \cos \theta - \alpha \sin^2 \theta - \frac{\alpha N^2}{N^2 - \alpha^2}, \\
\Delta &= r^2 - 2mr + N^2 - \alpha^2 + p^2 - q^2, & \Sigma &= P_r - \alpha P_\theta = r^2 - (N + \alpha \cos \theta)^2.
\end{align*}

This metric is a generalization of the Euclidean-Kerr-Taub-Bolt (EKTB) metric. Indeed if one takes $p^2=q^2$, one recovers the former. We will see in more details in section \ref{sec:self dual solution} how to recover the running-bolt solution from this one. Note that the metric depends on $p$ and $q$ only through $\delta \equiv q^2-p^2$.

Here we review the conditions that the regularity and smoothness of the four-dimensional metric impose on the choices of parameters. For more details, see \cite{Bobev:2009kn} and \cite{Bena:2009qv,bena2025positivespininducedquadrupolemoment} as the logic of the analysis is the same.

The metric has horizons provided the roots of $\Delta$, denoted $r_\pm$, are real, with:
\begin{equation}
    r_\pm \equiv m \pm \sqrt{m^2-N^2+\alpha^2+\delta}
\end{equation}

This requires the condition:
\begin{equation}
    m^2 \geq N^2-\alpha^2+\delta\, .
\end{equation}

To ensure the quantization of the NUT charge / avoid non-removable Dirac strings, the coordinates must satisfy the following periodicity condition \cite{Gibbons:1979nf,Bena:2009qv}:
\begin{equation}
\label{periodicity_cond_inf}
(\tau, \phi)\sim (\tau - 8N\pi, \phi + 2\pi)\, .
\end{equation}
Similarly, to prevent conical singularities at the outer horizon $r_+$, one can do a change of coordinates to analyze this near-horizon region. The analysis shows then that avoiding conical singularities requires another condition on the periodicity of $\phi$ and $\tau$ \cite{Gibbons:1979nf,Bena:2009qv}:
\begin{equation}
\label{periodicity_cond_rplus}
\left(\tau, \phi\right) \sim \Big( \tau - \frac{2 \pi}{\kappa},\phi + \frac{2 \pi \alpha }{\kappa \abs{P_{r_{+}}}}\Big),
\end{equation}
where we defined $\kappa \equiv \abs{ \frac{r_{+}-r_{-}}{2 P_{r_{+}}}}$.

For the geometry to be globally regular, these two periodicity lattices must be perfectly compatible, and the angular coordinate $\phi$ must remain strictly $2\pi$-periodic. Matching equations \eqref{periodicity_cond_inf} and \eqref{periodicity_cond_rplus} yields two constraints:
\begin{equation}
\frac{ \alpha}{\kappa \abs{P_{r_{+}}}} = \frac{ \alpha}{\sqrt{m^2 - N^2 +\alpha^2+\delta}}\equiv \ell \in {\mathbb Z}
\label{periodicity of phi}
\end{equation}
and
\begin{equation}
\label{N and kappa}
\frac{1}{\kappa} = 4  \abs{N}\, .
\end{equation}
Without loss of generality, we can assume that $\alpha$ and $\ell$ are positive, as one can always change the orientation of $\tau$ and the sign of $N$ to achieve this. The physical parameters $N,m,\alpha$ and $\delta$ are thus linked by two equations. Let us analyze the second one. Because $r_+$ involves a square root, enforcing \eqref{N and kappa} produces not two but four branches of solutions, depending on the relative signs of $N$ and $P_{r_+}$ \cite{Gibbons:1979nf,Bena:2009qv}.
When $N$ and $P_{r_+}$ have the same sign, regular solutions must satisfy:
\begin{multline}
\label{eq:Prplus and N same sign}
	16 N \left(N^2 - \alpha^2\right)^2 m^3 - 4\left(N^2 - \alpha^2\right)\left(5N^4 - 3N^2 \alpha^2 - \alpha^4\right)m^2  \\
	- 16 N \left(N^2 - \alpha^2\right)^2 \left(N^2 - \alpha^2 -\delta\right) m + 20 N^8 - 52 N^6 \alpha^2 + 49 N^4 \alpha^4 - 16 N^2 \alpha^6  \\
	- 2 N^2 \delta \left(N^2 - \alpha^2\right)\left(10 N^2 - 9 \alpha^2\right) + \delta^2 \left(N^2 - \alpha^2\right)^2 = 0
\end{multline}
Conversely, when $N$ and $P_{r_+}$ have opposite signs, the constraint becomes:
\begin{multline}
\label{eq:Prplus and N diff sign}
	-16 N \left(N^2 - \alpha^2\right)^2 m^3 - 4\left(N^2 - \alpha^2\right)\left(5N^4 - 3N^2 \alpha^2 - \alpha^4\right)m^2  \\
	+ 16 N \left(N^2 - \alpha^2\right)^2 \left(N^2 - \alpha^2 -\delta\right) m + 20 N^8 - 52 N^6 \alpha^2 + 49 N^4 \alpha^4 - 16 N^2 \alpha^6  \\
	- 2 N^2 \delta \left(N^2 - \alpha^2\right)\left(10 N^2 - 9 \alpha^2\right) + \delta^2 \left(N^2 - \alpha^2\right)^2 = 0
\end{multline}
When $\delta=0$, these equations reduce to the ones of the Kerr-Taub-Bolt base space \cite{bena2025positivespininducedquadrupolemoment}.

Since equations~\eqref{eq:Prplus and N same sign}-\eqref{eq:Prplus and N diff sign} are obtained by squaring the original regularity constraint~\eqref{N and kappa}, there is an extra inequality coming from requiring that the LHS and RHS of \eqref{N and kappa} have the same sign.
Furthermore, \eqref{N and kappa} is equivalent to $P_{r_+}=\pm 4N s$, with $s\equiv\sqrt{m^2-N^2+\alpha^2+\delta}$, where the upper (lower) sign corresponds to $N$ and $P_{r_+}$ having the same (opposite) sign. Multiplying by $N^2-\alpha^2$, this condition becomes
\begin{equation}
2\,(m\mp 2N)(N^2-\alpha^2)\, s + (N^2-\alpha^2)(2m^2+\delta) - N^2(2N^2-\alpha^2) =0 \, ,
\end{equation}
and squaring it gives \eqref{eq:Prplus and N same sign} or \eqref{eq:Prplus and N diff sign}. Since $s\geq 0$, the extra inequality is
\begin{equation}
\label{eq:extra inequality}
(m\mp 2N)(N^2-\alpha^2)\Big[(N^2-\alpha^2)(2m^2+\delta) - N^2(2N^2-\alpha^2)\Big] \leq 0 \, .
\end{equation}

As a result, we have a system of four free parameters and two equations. Since all equations are homogeneous of degree 8 in $m,N,\alpha,\delta$ (where $\delta$ counts with degree two), we can set $|N|$ to be $1$. However there are still too many free parameters and we do not have solution points as in \cite{bena2025positivespininducedquadrupolemoment}, but line solutions: we can express $\delta$ and $\alpha$ with respect to $m$. Then fixing $\delta(m)$ fixes $m$ and thus $\alpha(m)$, and we have a solution point $(m,\alpha,\delta)$.

\begin{figure*}[h!]
        \centering
        \includegraphics[width=0.6\linewidth]{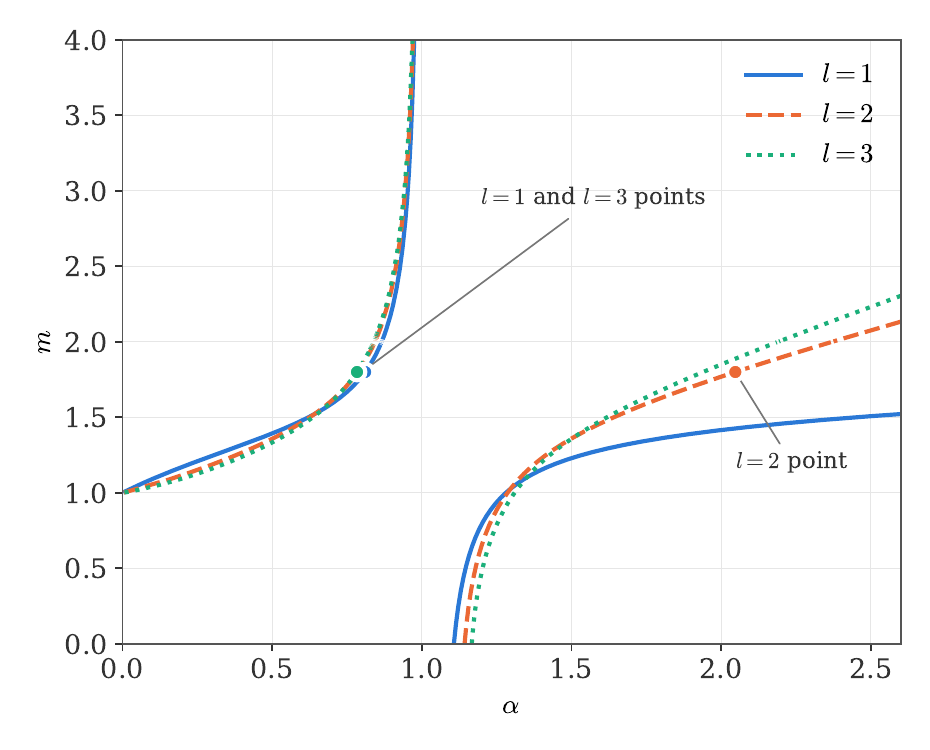}
        \caption{Base spaces with flat four-dimensional asymptotics, for $N=1$ and $N$ and $P_{r_+}$ of the same sign. Each curve gives the solutions $m(\alpha)$ of \eqref{eq:Prplus and N same sign} that satisfy \eqref{periodicity of phi} for $\ell=1$ (solid), $\ell=2$ (dashed) and $\ell=3$ (dotted), and the extra inequality \eqref{eq:extra inequality}. On a point on the curve $\delta$ is fixed by \eqref{periodicity of phi}, and varies continuously along the curve. The dots are the base spaces with $m=1.8$ used in Table~\ref{tab:results N=1 selfdual}.} 
        \label{fig:allowed regions m(alpha) for N>0}
\end{figure*}

Concretely, for a given $\ell$, equation \eqref{periodicity of phi} gives $\delta=\alpha^2/\ell^2-m^2+N^2-\alpha^2$, and \eqref{eq:Prplus and N same sign} or \eqref{eq:Prplus and N diff sign} then becomes a curve in the $(\alpha,m)$ plane. We represent these curves for $N=1$, $\ell=1,2,3$ and $N$ and $P_{r_+}$ of the same sign in Figure~\ref{fig:allowed regions m(alpha) for N>0}. All the solutions we study in Section~\ref{sec:regular} belong to this branch.

Note that equation \eqref{periodicity of phi} is already imposed in Figure~\ref{fig:allowed regions m(alpha) for N>0}: each curve corresponds to one value of $\ell$. In this paper we only study profiles with $\ell=1,2,3$ for more clarity; solutions with larger $\ell$ can be found in exactly the same way.

The base space \eqref{eq:BRbase} has signature $(+,+,+,+)$ everywhere outside the bolt if $\Sigma>0$ for $r>r_+$, that is, if $r_+>|N|+\alpha$. For the Kerr-Taub-Bolt base spaces with flat four-dimensional asymptotics this never happens: they always have $|N|\geq m$ and hence regions where the signature changes from $(+,+,+,+)$ to $(-,-,-,-)$ \cite{bena2025positivespininducedquadrupolemoment}. For the Bobev-Ruef base spaces the parameter $\delta$ allows $m>|N|$, and the condition can be satisfied: for example, the $\ell=1$ and $\ell=3$ base spaces of Table~\ref{tab:results N=1 selfdual} are positive definite everywhere, while the $\ell=2$ one is ambipolar. Anyway, we are interested in the regularity of the eleven-dimensional fluxed solution, and the four-dimensional base metric can be ambipolar as long as the full solution is not. Indeed, when fluxes are present, the warp factors and rotation parameter blow up in exactly the precise way to give rise to a regular five-dimensional solution \cite{Bena:2009qv}.

We note that the periodicity relation \eqref{periodicity of phi} breaks down in the extremal limit, when $m^2 - N^2 +\alpha^2+\delta = 0$ and the horizons degenerate to $r_+=r_- =m$.
This regime needs to be studied separately; we come back to it in Section~\ref{sec:BPS limit}.

\subsection{Adding fluxes}\label{sec:fluxes}

Following the work done in \cite{Bobev:2009kn,Bena:2009fi}, we recall the results of their section 2. One can construct from this four-dimensional base space an eleven-dimensional supergravity solution given by:

\begin{equation}
\label{eq:11Dmetric}
\begin{split}
    ds_{11}^2=&-\left(Z_1 Z_2 Z_3\right)^{-2/3}(dt+k)^2+\left(Z_1 Z_2 Z_3\right)^{1/3}ds_4^2\\
    &+\left(Z_1 Z_2 Z_3\right)^{1/3} \left(\dfrac{ds^2\left(T_1^2\right)}{Z_1}+\dfrac{ds^2\left(T_2^2\right)}{Z_2}+\dfrac{ds^2\left(T_3^2\right)}{Z_3}\right),
\end{split}
\end{equation}
where the three two-tori are wrapped by the magnetic fluxes $\Theta^{(I)}$. The conventions used for eleven-dimensional supergravity are recalled in Appendix~\ref{Appendix Supergravity}.

The equations of motion are \cite{Bena:2009fi}:

\begin{align}
\hat{\nabla}^2 Z_1 &= *_4 \left( \Theta^{(2)} \wedge \Theta^{(3)} \right), & \left( \Theta^{(2)} - *_4 \Theta^{(2)} \right) &= 2 Z_1 \omega_{-}^{(3)}, \\
\hat{\nabla}^2 Z_2 &= *_4 \left( \Theta^{(1)} \wedge \Theta^{(3)} \right), & \left( \Theta^{(1)} - *_4 \Theta^{(1)} \right) &= 2 Z_2 \omega_{-}^{(3)},
\end{align}
\begin{align}
     \hspace{1cm} \hat{\nabla}^2 Z_3 &= *_4 \left[ \Theta^{(1)} \wedge \Theta^{(2)} - \omega_{-}^{(3)} \wedge \left( dk - *_4 dk \right) \right], \\
 dk + *_4 dk &= \frac{1}{2} \sum_{I=1}^{3} Z_I \left( \Theta^{(I)} + *_4 \Theta^{(I)} \right).
\end{align}
Here $\hat\nabla^2$ and $*_4$ are the Laplacian and the Hodge dual of the base \eqref{eq:BRbase}, and $\Theta^{(3)}$ and $\omega_-^{(3)}$ are respectively the self-dual and the anti-self-dual parts of the Maxwell field of the base. When the Maxwell field is self-dual, $\omega_-^{(3)}=0$, and these equations reduce to the ones used to construct the Running-Bolt solutions \cite{Bena:2009qv,bena2025positivespininducedquadrupolemoment}.

One important remark is that the expression of $Z_3$ will be different from $Z_1$ and $Z_2$ and therefore when charges will be computed, one should not treat the three tori equivalently.

\subsection{Self-dual solution}\label{sec:self dual solution}

The fluxes that are solutions of the equations of the previous section are \cite{Bobev:2009kn}:

\begin{equation}
\label{eq:selfdualfluxes}
     \left\{
    \begin{array}{ll}
        \Theta^{(1)} &= f_1\left(r, \theta\right) \Omega_+ + g_1\left(r, \theta\right) \Omega_-,\\[1em]
        \Theta^{(2)} &= f_2\left(r, \theta\right) \Omega_+ + g_2\left(r, \theta\right) \Omega_-,\\[1em]
        \Theta^{(3)} &= \left(p+q\right) dA_+,
    \end{array}
\right. \, .
\end{equation}

where we used:
\begin{align}
A_{\pm} &= - \frac{1}{r \mp \left(N + \alpha \cos\theta\right)} \left(d\tau + P_{\theta} d\phi\right) \mp \cos\theta d\phi,\\
f_2 &= \frac{2q_2}{\left[r - \left(N + \alpha \cos\theta\right)\right]^2} - \frac{2q_2\left(p^2 - q^2\right)}{m-N} \frac{1}{\left[r - \left(N + \alpha \cos\theta\right)\right]^2 \left[r + \left(N + \alpha \cos\theta\right)\right]},  \\
g_2 &= \frac{p-q}{\left[r + \left(N + \alpha \cos\theta\right)\right]^2} - \frac{2q_2\left(p^2 - q^2\right)}{m-N} \frac{1}{\left[r - \left(N + \alpha \cos\theta\right)\right] \left[r + \left(N + \alpha \cos\theta\right)\right]^2}, \\
f_1 &= \frac{2q_1}{\left[r - \left(N + \alpha \cos\theta\right)\right]^2} - \frac{2q_1\left(p^2 - q^2\right)}{m-N} \frac{1}{\left[r - \left(N + \alpha \cos\theta\right)\right]^2 \left[r + \left(N + \alpha \cos\theta\right)\right]},  \\
g_1 &= \frac{p-q}{\left[r + \left(N + \alpha \cos\theta\right)\right]^2} - \frac{2q_1\left(p^2 - q^2\right)}{m-N} \frac{1}{\left[r - \left(N + \alpha \cos\theta\right)\right] \left[r + \left(N + \alpha \cos\theta\right)\right]^2},\\
\Omega_\pm &=dr \wedge \left(d\tau+ P_\theta d\phi\right)\pm \sin\theta d\theta \wedge \left(\alpha d\tau + P_r d\phi\right).
\end{align}
One can check that $dA_\pm=\Omega_\pm/\left[r\mp(N+\alpha\cos\theta)\right]^2$, and that the anti-self-dual two-form entering the equations of motion is $\omega_-^{(3)}=(p-q)\,dA_-$. In particular, $g_1=(p-q)Z_2/\left[r+(N+\alpha\cos\theta)\right]^2$ and $g_2=(p-q)Z_1/\left[r+(N+\alpha\cos\theta)\right]^2$, with $Z_1$ and $Z_2$ given below.

The warp-factors and rotation vector solving the equations are:
\begin{equation}
\label{eq:wrapfactors}
     \left\{
    \begin{array}{ll}
        Z_1 &= 1 - \dfrac{2q_2\left(p+q\right)}{m-N} \dfrac{1}{r - \left(N + \alpha \cos\theta\right)},\\[1em]
        Z_2 &= 1 - \dfrac{2q_1\left(p+q\right)}{m-N} \dfrac{1}{r - \left(N + \alpha \cos\theta\right)}, \\[1em]
        Z_3 &= 1 - \dfrac{4q_1 q_2}{\left(m-N\right)} \dfrac{1}{r - \left(N + \alpha \cos\theta\right)} + \dfrac{4q_1 q_2 \left(p^2 - q^2\right)}{\left(m-N\right)^2} \dfrac{1}{\Sigma} \nonumber \\
&\quad + \left( \dfrac{4q_1 q_2 \left(p^2 - q^2\right)}{\left(m-N\right)^3} - \dfrac{2\left(q + q_1 + q_2\right)\left(p-q\right)}{m-N} + \lambda\left(m-N\right) \right) \dfrac{1}{r + \left(N + \alpha \cos\theta\right)},\\[1 em]
        k &= \dfrac{1}{\Sigma} \left( \left(\Sigma \mu - \alpha \nu\right)\left(d\tau + P_{\theta} d\phi\right) + \nu\left(\alpha d\tau + P_r d\phi\right) \right) = \mu\left(d\tau+P_\theta d\phi\right)+\nu\, d\phi.
    \end{array}
\right.
\end{equation}
The scalars appearing in $Z_3$ and $k$ are $\mu,\nu$ and $\lambda$, given by (see equations (4.42), (4.44) and (4.45) of \cite{Bobev:2009kn}):

\begin{align}
\lambda &= \frac{4N\left(p-q\right)}{p^2 - q^2 - 2\left(m+N\right)r_+} \left( \frac{2q_1 q_2\left(p+q\right)\left(m+N\right)}{\left(m-N\right)^4} - \frac{\left(q_1 + q_2\right)\left(p^2 - q^2\right)}{\left(m-N\right)^3} + \frac{\left(p-q\right)r_+}{\left(m-N\right)^2} \right) \label{eq:lambda} \\
\nu &= \left( \gamma - \frac{4q_1 q_2\left(p+q\right)}{\left(m-N\right)^2} \frac{1}{r - \left(N + \alpha \cos\theta\right)} \right) \alpha \sin^2\theta, \label{eq:nu} \\
\gamma &= \frac{q+q_1+q_2}{m-N} - \frac{2q_1 q_2\left(p+q\right)}{\left(m-N\right)^3} - \lambda \frac{m^2 - N^2}{4N\left(p-q\right)}\,, \label{eq:gamma}
\end{align}
and $\mu$ is given below. The constant $\gamma$ is the asymptotic value of $\mu$. Note that, as for the Running-Kerr-Taub-Bolt solutions, $\gamma$ depends on the position of the bolt, $r_+$, through $\lambda$.

Note that there was a typo in \cite{Bobev:2009kn} for the expression of $\mu$, which is:
\begin{equation}
\label{eq:mu}
    \begin{split}
        \mu=&\frac{1}{\Sigma }\Bigg(\frac{1}{2} \lambda (N-m) (p+q)\\
        &+\frac{2 q_1 q_2 (p+q) \left(-2 \alpha ^2-m (N+3 r)-\alpha  \cos (\theta ) (m-r)+2 N^2+N r+2 p^2-2 q^2+r^2\right)}{(m-N)^2 (\alpha  \cos (\theta )+N-r)}\\
        &-\frac{\left(p^2-q^2\right) \left(2 q_1 q_2 (p+q)-(m-N)^2 (q+2 (q_1+q_2))\right)}{(m-N)^3}+\frac{\alpha ^2 \sin ^2(\theta ) (4 p q_1 q_2+4 q q_1 q_2)}{(m-N)^2 (\alpha  \cos (\theta )+N-r)}\\
        &-\frac{(2 (N-r)) (p+q) (q_1 q_2)}{(m-N)^2}+\gamma  \left((r-N)^2-\alpha ^2 \cos ^2(\theta )\right)-r (p+q+2 (q_1+q_2))\Bigg).
    \end{split}
\end{equation}

Note that from the expression of the warp factors and as noted in \cite{Bobev:2009kn}, we explicitly see the limit to recover the Running-Kerr-Taub-Bolt solution \cite{Bena:2009qv,bena2025positivespininducedquadrupolemoment}. One needs to take $p=q$ and to rescale:
\begin{equation}
\label{eq:correspondence KTB BR}
	(q_{1,KTB},q_{2,KTB},q_{3,KTB})=(2q_{1,BR},2q_{2,BR},2p_{BR})\, .
\end{equation}

This limit will be useful for us to recover the KTB limit in our results.

From this eleven-dimensional solution, one can do a Kaluza-Klein reduction to ten dimensions. However to do so, one needs to be careful because in this system of coordinates the asymptotic spacetime is not static and therefore one cannot do a Kaluza-Klein reduction directly, but needs to do a change of coordinates to end up in an eleven dimensional metric which is asymptotically at rest. This change of variables is:

\begin{equation}
     \left\{
    \begin{array}{ll}
    \hat{\tau} = (1 - \gamma^2)^{1/2} (\tau - \dfrac{\gamma}{1-\gamma^2} t)\\
    \hat{t} = (1 - \gamma^2)^{-1/2} t
    \end{array}
\right. \,\ \  .
\end{equation}
This is a well-defined coordinate transformation if the bolt runs slower than the speed of light, $|\gamma|<1$.

Performing this change of variables we rewrite the metric in a form ready for Kaluza-Klein reduction:

\begin{align}
ds_5^2 &= \frac{\hat{I}_4}{\left(ZV\right)^2} \left( d\hat{\tau} + \left( \gamma - \frac{\mu V^2}{\hat{I}_4} \right) d\hat{t} + \left( \hat{P}_{\theta} - \frac{\hat{\nu} \mu V^2}{\hat{I}_4} \right) d\phi \right)^2 + \frac{ZV}{\hat{I}_4^{1/2}} ds_E^2,  \\
ds_E^2 &= -\hat{I}_4^{-1/2} \left( d\hat{t} + \hat{\nu} d\phi \right)^2 + \hat{I}_4^{1/2} \left( \frac{\Delta_{\theta}}{\Delta} dr^2 + \Delta_{\theta} d\theta^2 + \Delta \sin^2\theta d\phi^2 \right), \label{eq:dsE}
\end{align}

with:

\begin{align}
    \hat{I}_4 &= \left(1-\gamma^2\right)^{-1} \left(Z_1 Z_2 Z_3 V - \mu^2 V^2\right), \\
    \hat{P}_{\theta} &= \left(1-\gamma^2\right)^{1/2} \tilde{P}_{\theta},\\
    \hat{\nu} &= \left(1-\gamma^2\right)^{-1/2} \left(\nu - \alpha \dfrac{\Sigma}{\Delta_{\theta}} \sin^2\theta \, \mu\right),
\end{align}
where $Z\equiv(Z_1Z_2Z_3)^{1/3}$, and where we wrote the base metric \eqref{eq:BRbase} in the form
\begin{equation}
ds_4^2 = V^{-1}\left(d\tau+\tilde P_\theta d\phi\right)^2 + V\left(\frac{\Delta_\theta}{\Delta}dr^2+\Delta_\theta d\theta^2+\Delta\sin^2\theta d\phi^2\right),
\end{equation}
with
\begin{equation}
\Delta_\theta\equiv\Delta+\alpha^2\sin^2\theta\,,\qquad V\equiv\frac{\Sigma}{\Delta_\theta}\,,\qquad \tilde P_\theta \equiv P_\theta+\alpha\,\frac{\Sigma}{\Delta_\theta}\sin^2\theta\, .
\end{equation}
The metric $ds_E^2$ is the four-dimensional Einstein-frame metric.
Physical consistency requires that this four-dimensional metric be positive definite and free of closed timelike curves (CTCs). Concretely, we require that everywhere outside the bolt, $r>r_+$,
\begin{equation}
\label{eq:CTC conditions}
Z_I V>0\,,\qquad \hat I_4>0\,,\qquad \hat I_4\,\Delta\sin^2\theta-\hat\nu^2>0\,,
\end{equation}
and that the bolt runs slower than the speed of light, $|\gamma|<1$. These requirements are far from trivial: they hugely restrict the admissible values of the fluxes $q_I$ and $p$.
The resulting constraints are collected in Table~\ref{tab:results N=1 selfdual}. In this paper we focus on solutions with $N>0$ and $m>0$. Using the homogeneity of the equations, we can then take $N=1$ without any loss of generality and compute the multipole moments for the allowed $(m',\alpha')$.
Unlike for the RKTB solution, a numerical scan with the conditions \eqref{eq:CTC conditions} finds CTC-free solutions with $N<0$ (and $m>0$), and with $N>0$, $m<0$ (e.g. $N=1$, $m=-0.6$, $\ell=1$), all with $M>0$.

\section{Multipole moments and charges}
\label{sec:multipoles}

To extract the physical asymptotic observables of our four-dimensional geometry—namely the mass, angular momentum, and higher-order multipole moments—one can rely on either the Geroch-Hansen formalism \cite{Geroch:1970cd,Hansen:1974zz} or Thorne's multipole expansion \cite{RevModPhys.52.299}. As these two approaches are equivalent \cite{Gursel:1983nkl}, we will frame our computation using the latter.

Following Thorne's method for stationary, axisymmetric spacetimes, one must perform an appropriate coordinate transformation at spatial infinity to reach an ``Asymptotically Cartesian to order N'' (AC-N) frame. By applying a further spatial shift to place the center of mass exactly at the origin, this frame is promoted to an ``Asymptotically Cartesian and Mass Centered to order N'' (ACMC-N) coordinate system. The defining feature of this mass-centered frame is the strict vanishing of the mass dipole moment, $M_1=0$. The precise asymptotic form such a coordinate system must have is very constrained and detailed, for instance, in \cite{Bena:2020uup}.
Once the geometry is cast in this properly adapted frame, the asymptotic mass and the gravitational multipoles can be systematically read off from the expansion of the metric at infinity. In the following analysis, we will use $\tilde M_n$ to denote the ``fake'' multipole moments extracted in a generic, non-centered AC-N frame, while $M_n$ will stand for the ``true'' physical multipole moments evaluated strictly in the proper ACMC-N coordinate system.

Expanding the four-dimensional metric \eqref{eq:dsE} at spatial infinity reveals that our initial coordinate system does not naturally satisfy the AC-N fall-off conditions. Guided by similar procedures in the literature \cite{Bena:2020uup,Heidmann:2025yzd,Bah:2022yji,bena2025positivespininducedquadrupolemoment}, we introduce a new set of coordinates $(r_s,\theta_s)$, which are prolate spheroidal coordinates shifted by $b$ along the axis:
\begin{equation}
    \left\{
    \begin{array}{ll}
    r_s \sin \theta_s = \sqrt{r^2-\alpha^2}\sin \theta \\
    r_s \cos \theta_s -b = r \cos \theta
    \end{array}
    \right. \, \ .
\end{equation}

By expanding the metric in these new coordinates, one can verify that the system becomes AC-N for every order of N, it is thus  AC-$\infty$.

The constant $b$ accounts for the displacement of the center of mass, which corresponds to a simple shift along the $z$-axis in cylindrical coordinates. To make our coordinate system fully mass-centered (ACMC-$\infty$), we fix the shift, $b$, by requiring the mass dipole moment, $M_1$, to vanish exactly.

In this frame, we can read the mass, the angular momentum, and the mass dipole and quadrupole moment $M_2$  directly from the asymptotic expansions of $g_{\hat{t}\hat{t}} $ and $g_{\hat{t}\phi} $ (higher multipole moments can be evaluated just as easily). Indeed in AC-$\infty$ coordinates these metric components are related to the physical quantities via:
\begin{equation}
    \begin{split}
        g_{\hat{t}\hat{t}} &= -1+\frac{2G_4M}{r_s}+ \frac{2\left(G_4\tilde M_1 \cos\theta_s+c_1\right)}{r_s^2}+ \frac{(3 \cos^2{\theta_s}-1)G_4\tilde{M_2}+ c_2 \cos{\theta_s}+c_3}{r_s^3} +O\left(\frac{1}{r_s^4}\right)\\
        g_{\hat{t}\phi} &= -2 \sin^2\theta_s \frac{G_4 S_1}{r_s} + O\left(\frac{1}{r_s^2}\right)
    \end{split}
\end{equation}
Here, $c_i$ are constants, and $\tilde M_1$ and $\tilde M_2$ are the ``fake'' multipole moments, as they are measured in a frame that is not yet mass-centered.
$\tilde M_1$ is thus obtained in terms of $b$. We can now fix this parameter by requiring $\tilde M_1$ to vanish. For simplicity, we give the result when the fluxes on the first two tori are equal, $q_1=q_2$, which is the case we study in the rest of this paper:
\begin{equation}
\label{eq:b}
\begin{split}
    b = \frac{\alpha}{D_b}  \Big(&\lambda (N-m)^4-4 \delta  q_1^2\\
    &+2 (N-m) \left((N-m) \left(-N (N-m)-p q+q^2+4 q q_1+2 q_1^2\right)+4 \gamma  q_1^2 (p+q)\right)\Big)\ ,
\end{split}
\end{equation}
where
\begin{equation}
	\begin{split}
    D_b &= m^4 \left(
    -4 \gamma ^2+\lambda+2\right)+2 m^3 \left(N \left(8 \gamma ^2-2 \lambda-3\right)+\gamma  (p+q+4 q_1)\right)\\
    & \quad +2 m^2 \left(3 N \left(N \left(-4 \gamma ^2+\lambda+1\right)-\gamma  (p+q+4 q_1)\right)-p (q+4 q_1)+q^2-2 q_1^2\right)\\
    & \qquad +2 m N \left(N^2 \left(8 \gamma ^2-2 \lambda-1\right)+3 \gamma  N (p+q+4 q_1)+2 q (p-q)+8 p q_1+4 q_1^2\right) \\	& \qquad \quad +N^2 \left(N (\lambda N-2 \gamma  (2 \gamma  N+p+q+4 q_1))-2 \left(q (p-q)+4 p q_1+2 q_1^2\right)\right)-4 \delta  q_1^2 \ .
    \end{split}
\end{equation}
One then obtains:

\begin{align}
    G_4 M &= \frac{1}{4(1-\gamma^2)} \Bigg[ 2m + \frac{4q_1^2 + 4q_1(p+q) + 2(p-q)(q+2q_1)}{N-m} \nonumber \\
    &\qquad\qquad\qquad\quad + \frac{4\delta q_1^2}{(N-m)^3} - \lambda(N-m) + 2\gamma(p+q+4q_1) + 4\gamma^2(N-m) \Bigg] \, , \label{M self dual} \\[1em]
    G_4 S_1 &= G_4 J = -\frac{\alpha}{2\sqrt{1-\gamma^2}} \left[ \frac{4q_1^2(p+q)}{(N-m)^2} - 2\gamma(N-m) - p - q - 4q_1 \right] \, , \label{J self dual}\\[1em]
\begin{split}
    G_4 M_2 &= \frac{\alpha^2}{4(1-\gamma^2)(m-N)^6} \Bigg\{ (m-N)^3 \bigg( m^4(-4\gamma^2+\lambda+2) \\
    &\quad + 2m^3\big[N(8\gamma^2-2\lambda-3) + \gamma(p+q+4q_1)\big] \\
    &\quad + 2m^2\big[3N\big(N(-4\gamma^2+\lambda+1)-\gamma(p+q+4q_1)\big) - p(q+4q_1) + q^2 - 2q_1^2\big] \\
    &\quad + 2mN\big[N^2(8\gamma^2-2\lambda-1) + 3\gamma N(p+q+4q_1) + 2q(p-q) + 8pq_1 + 4q_1^2\big] \\
    &\quad + N^2\big[N(\lambda N - 2\gamma(2\gamma N+p+q+4q_1)) - 2(q(p-q)+4pq_1+2q_1^2)\big] - 4\delta q_1^2 \bigg) \\
    &\quad - \frac{1}{\mathcal{D}_{M_2}} \bigg( \lambda(m-N)^4 + 2(m-N)\big[(m-N)(N(m-N)-pq+q^2+4qq_1+2q_1^2) \\
    &\qquad\qquad\qquad\qquad - 4\gamma q_1^2(p+q)\big] - 4\delta q_1^2 \bigg)^2 \Bigg\} \, ,
\end{split} \label{M J M2 self dual}
\end{align}
where the denominator of the second term in $G_4 M_2$ is defined as:

\begin{equation}
\begin{split}
\mathcal{D}_{M_2} &= 2m+\left(m-N\right) \left(\lambda-4 \gamma ^2\right) \\[0.5em]
	&\qquad+ 2 \left(p+q+4 q_1\right) \gamma-\frac{2\left(pq-q^2+4pq_1+2q_1^2\right)}{m-N}-\frac{4 q_1^2\delta}{(m-N)^3}\, .
\end{split}
\end{equation}
These expressions simplify drastically. Comparing with \eqref{M self dual}, one finds that
\begin{equation}
\mathcal{D}_{M_2} = 4(1-\gamma^2)\,G_4M\,,\qquad D_b = (m-N)^3\,\mathcal{D}_{M_2}\,,
\end{equation}
and that the second term in \eqref{M J M2 self dual} is precisely $b\,D_b/\alpha$. Hence the quadrupole moment takes the remarkably simple form
\begin{equation}
\label{eq:M2 simple}
	M_2 = M\,\left(\alpha^2-b^2\right)\,.
\end{equation}
Since the mass of the regular solutions is positive, the quadrupole moment is positive if and only if $|b|<\alpha$. One can check that the same relation holds for the Running-Kerr-Taub-Bolt solutions of \cite{bena2025positivespininducedquadrupolemoment}. Note that, if the center of mass were matching the center of the bolt ($b=0$), the quadrupole moment would be $M_2=M\alpha^2$, which is the analytic continuation of the Kerr relation $M_2=-Ma^2$ with the Euclidean rotation parameter, $a^2\to-\alpha^2$. This displacement of the center of mass decreases $M_2$, and is due to the asymmetry of the geometry. When the center of mass is not translated too much (less than $\alpha$) the multipole remains positive.

Our aim is to evaluate the sign of $G_4M_2$ for smooth solutions with four-dimensional asymptotics. All the quantities above are homogeneous functions of the parameters $(m,N,\alpha,p,q,q_1,q_2)$ (with $\lambda$ and $\gamma$ of degree zero): $G_4M$, $G_4J$ and $G_4M_2$ are of degree one, two and three respectively. We therefore introduce the homogeneous parameters
\begin{align*}
        m' &\equiv \frac{m}{\abs{N}}\ ,  & \alpha' &\equiv \frac{\alpha}{\abs{N}}\ , & \delta' &\equiv \frac{\delta}{N^2}\ , &
        p' &\equiv \frac{p}{\abs{N}}\ , & q' &\equiv \frac{q}{\abs{N}}\ , & q_I' &\equiv \frac{q_I}{\abs{N}}\ ,
\end{align*}
so that, for example, $ M_2 (m,\alpha,N,p,q,q_I) = \abs{N}^3 M_2 \big(m',\alpha',N/|N|,p',q',q_I'\big)$. The dimensionless multipole ratio
\begin{equation}
R\equiv\frac{M_2M}{J^2}\,,
\end{equation}
introduced in \cite{Bena:2020see}, does not depend on $|N|$. The Kerr black hole has $R=-1$.

\section{Regular solutions}
\label{sec:regular}

For a given base space, that is for given $(m',\alpha',\delta')$ satisfying the regularity conditions of Section~\ref{sec:base}, the solutions are determined by the fluxes $p$, $q$, $q_1$ and $q_2$, with $q^2=p^2+\delta$. Even with $q_1=q_2$, this leaves a two-dimensional space of solutions for each base space, and the conditions \eqref{eq:CTC conditions} must be checked at every point $(r,\theta)$ outside the bolt. It is therefore not possible to plot all the solutions, and we proceed as follows:
\begin{enumerate}
\item We fix $\ell$ and $m'$. The regularity conditions \eqref{periodicity of phi} and \eqref{eq:Prplus and N same sign} then fix $\alpha'$ and $\delta'$ (Figure~\ref{fig:allowed regions m(alpha) for N>0}), and we check the inequality \eqref{eq:extra inequality} This step corresponds to choosing a point of the curve in Figure \ref{fig:allowed regions m(alpha) for N>0}, for a given $\ell$.
\item We study the condition that does not depend on other variables, namely $|\gamma|<1$. After the first step, $\gamma$ depends only on two remaining variables: $p'$ and $q_{1}'$. Indeed we can substitute $q'$ by its expression with respect to $p'$ and $\delta'$ , $q'^2=p'^2+\delta'$. We choose $p'$ such that $|\gamma|<1$ for a large enough interval of $q_{1}'$.
\item We scan the remaining flux, $q_1'=q_2'$, and determine numerically the range where the conditions \eqref{eq:CTC conditions} are satisfied everywhere outside the bolt.
\item We compute $M_2$ and $R$ in this range.
\end{enumerate}
Each choice of $(\ell,m',p')$ thus gives one line of solutions in the two-dimensional space of solutions we have from fixing the base space.

The solutions have a simple symmetry: changing the signs of all the fluxes, $(p,q,q_1,q_2)\to-(p,q,q_1,q_2)$, leaves $\lambda$ invariant and flips the sign of $\gamma$. Under this transformation, $M$, $b$ and $M_2$ are invariant, $J$ changes sign, and $R$ is invariant. Hence, without loss of generality we can take $q>0$.

The results for $m'=1.8$ and $\ell=1,2,3$ are summarized in Table~\ref{tab:results N=1 selfdual}.

\begin{table}[h!]
    \centering
    \mbox{}\clap{
    \setlength{\tabcolsep}{3pt}{
    \begin{tabular}{|c|c|c|c|c|c|c|c|c|}
        \hline
        $\ell$ & \textbf{$m'$} & \textbf{$\alpha'$} & \textbf{$p'$} & $q'$ & \textbf{range of $q_1'$} & \textbf{$M_2$} & \textbf{ $M_2 >0$} & \textbf{$R\equiv{M_2 M}/{J^2}$} \\
        \hline
        1 & 1.8& 0.810578& 2.5 & 2.0025 & $-0.71\!<\!q_1'\!<\!-0.47$&$-\infty\!<\!M_2\!<\!3.28$ & $-0.69\!<\!q_1'\!<\!-0.53$ & $-\infty\!<\!R\!<\!0.93$\\
        \hline
        2 & 1.8&2.04738 & 2.5 & 0.9307 & $-1.39\!<\!q_1'\!<\!-1.07$& $-\infty\!<\!M_2\!<\!38.53$&$ -1.31\!<\!q_1'\!<\!-1.14$&$-\infty\!<\!R\!<\!0.29$ \\
        \hline
        3 & 1.8& 0.783504&2.3 & 1.5825 & $-1.59\!<\!q_1'\!<\!-1.09$& $-\infty\!<\!M_2\!<\!5.13$&  $-1.45\!<\!q_1'\!<\!-1.28$ & $-\infty\!<\!R\!<\!0.12$ \\
        \hline
        1 & 1.8 & 0.810578 & $-2.5$ & 2.0025 & $-0.24\!<\!q_1'\!<\!1.08$ & $-\infty\!<\!M_2\!<\!0.42$ & $-0.14\!<\!q_1'\!<\!0.74$ & $-\infty\!<\!R\!<\!2.07$ \\
        \hline

    \end{tabular}
    }
    }
    \caption{Regular solutions with $N=1$, $m'=1.8$, $\ell=1,2,3$ and $q_1=q_2$. For each base space ($\delta'=-2.24$, $-5.384$ and $-2.786$ for $\ell=1,2,3$), we fix $p'$, which fixes $q'=\sqrt{p'^2+\delta'}$, and give the range of $q_1'$ where the solution is free of CTCs, the range of the quadrupole moment, the range of $q_1'$ where it is positive, and the range of the multipole ratio $R$. The last line has the same base space as the first one, but the opposite sign of $p'$.}
    \label{tab:results N=1 selfdual}
\end{table}

\begin{figure*}[h!]
    \centering
    \begin{minipage}[t]{0.49\textwidth}
        \centering
        \includegraphics[width=\linewidth]{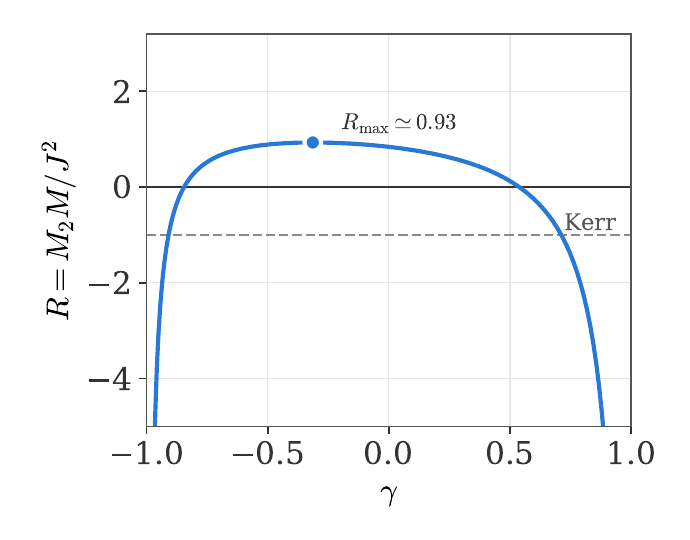}
        \caption{Plot of $R$ with respect to $\gamma$, for the solutions with $(m',\alpha',p')=(1.8,0.810578,2.5)$ ($\ell=1$) in the range of $q_1'$ free of CTCs. The ratio is bounded from above, reaches its maximum $R\simeq0.93$ at $\gamma\simeq-0.31$, and diverges when the boost becomes infinite, $|{\gamma}|\rightarrow 1$. The dashed line is the Kerr value $R=-1$.}
        \label{fig:multipole ratio R (M2) bounded}
    \end{minipage}
    \hfill
    \begin{minipage}[t]{0.49\textwidth}
        \centering
        \includegraphics[width=\linewidth]{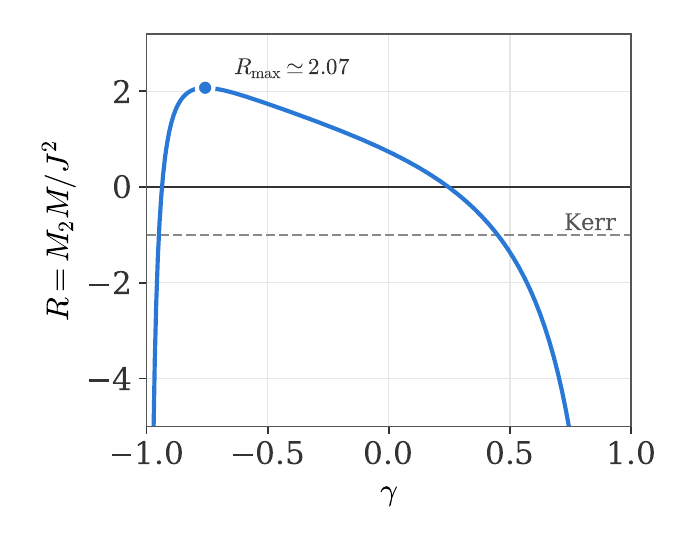}
        \caption{Plot of $R$ with respect to $\gamma$, for the solutions with $(m',\alpha',p')=(1.8,0.810578,-2.5)$ ($\ell=1$) in the range of $q_1'$ free of CTCs. The ratio is also bounded from above, reaches $R\simeq2.07$ at $\gamma\simeq-0.76$, and diverges when the boost becomes infinite, $|{\gamma}|\rightarrow 1$.}
        \label{fig:multipole ratio R (M2) unbounded}
    \end{minipage}
\end{figure*}

One finds that the spin-induced quadrupole moment, $M_2$, is positive within a finite interval of $q_1$, which always includes the value of $q_1$ for which the solution does not run ($\gamma=0$). As one increases the running speed, $|\gamma|$, the quadrupole moment flips sign and becomes negative. As the speed of the running bolt approaches the speed of light ($\gamma\to\pm1$), the mass and the quadrupole moment diverge and $R\to-\infty$. In all the examples we studied, the range of $q_1$ free of CTCs ends precisely where $|\gamma|\to 1$. An illustration of this behavior is shown in Figures~\ref{fig:multipole ratio R (M2) bounded} and~\ref{fig:multipole ratio R (M2) unbounded}.

There are two important differences with the Running-Kerr-Taub-Bolt solutions \cite{bena2025positivespininducedquadrupolemoment}. First, the maximum of $R$ is not necessarily reached at ``small'' velocities. Second, $R$ can be much larger: for the last line of Table~\ref{tab:results N=1 selfdual}, it reaches $R\simeq 2.07$ at $\gamma\simeq-0.76$, which is six times larger than the largest value we found for the Running-Kerr-Taub-Bolt solutions with $\ell=1,2,3$. In this example $M_2>J^2/M$: the quadrupole moment is larger in absolute value than that of a Kerr black hole with the same mass and angular momentum, but has the opposite sign.

Finally, as explained in Section~\ref{sec:base}, the base spaces of the first, third and fourth lines of Table~\ref{tab:results N=1 selfdual} have no region of $(-,-,-,-)$ signature. Hence, unlike for the Running-Kerr-Taub-Bolt solutions, a positive spin-induced quadrupole moment does not require an ambipolar base space.

\section{Page charges}
\label{sec:page}

\subsection{Definitions}

Brane charges can be defined in three different ways:
the \emph{Page charge}, which is conserved, localised and quantised, but invariant only under small gauge transformations;
the \emph{Maxwell charge}, which is gauge-invariant and conserved, but neither localised nor quantized;
and the \emph{brane-source charge}, which is gauge-invariant and localised, but neither conserved nor quantized \cite{pagecharge}.
Since these properties complement one another, the choice depends on the problem at hand. Here we use the Page charge: because it is conserved and quantized, it can be evaluated at spatial infinity, where the fields have a simple asymptotic form.

The charges of our solutions come from M$5$ and M$2$ branes. Reducing from $11D$ to $10D$ turns them into D$0$–D$2$–D$4$–D$6$ charges, and the subsequent reduction from $10D$ to $4D$ leaves their values unchanged. We can therefore compute the charges in $10D$, starting from the eleven-dimensional fields.

The reduction from $11D$ to $10D$ is a Kaluza–Klein (KK) reduction along the circle $\hat\tau$. Writing the eleven-dimensional metric and three-form in the standard form
\begin{equation}
\label{eq:KKform}
\begin{split}
    ds_{11}^2 &= E\,\bigl(d\hat\tau + C_1\bigr)^2 + E^{-1/2}\,ds_{10,E}^2\,,\\
    A_3 &= C_3 + B_2\wedge d\hat\tau\,,
\end{split}
\end{equation}
defines the ten-dimensional fields $C_1$, $B_2$ and $C_3$ in the string frame.

In the ten-dimensional frame, the Page charges and their normalisation in terms of brane numbers are given by \cite{de_Boer_2013,Chowdhury_2014} (see \cite{BenaLochetPage} for the derivation of the normalisation factors):

\begin{align}
    N_6^{\,p} &= \frac{1}{2\pi R_{\hat\tau}}
        \int \Bigl( dC_1 - B_2 \Bigr)\,,
    \label{eq:N6page}\\[6pt]
    N_4^{\,p} &= \frac{1}{(2\pi l_p)^3}
        \int \Bigl[
            dC_3 - dB_2\wedge C_1
            - B_2\wedge dC_1
            + \tfrac{1}{2}\,B_2\wedge B_2
        \Bigr]\,,
    \label{eq:N4page}\\[6pt]
    N_2^{\,p} &= \frac{2\pi R_{\hat\tau}}{(2\pi l_p)^6}
        \int \Bigl[
            {*\!\left(dC_3 - dB_2\wedge C_1\right)}
            - B_2\wedge\!\left(dC_3 - dB_2\wedge C_1\right)
            + \tfrac{1}{2}\,B_2\wedge B_2\wedge dC_1
        \Bigr]\,,
    \label{eq:N2page}\\[6pt]
    N_0^{\,p} &= \frac{(2\pi R_{\hat\tau})^2}{(2\pi l_p)^9}
        \int \Bigl[
            -\,{*\,dC_1}
            - B_2\wedge{*\!\left(dC_3 - dB_2\wedge C_1\right)}
            + \tfrac{1}{2}\,B_2\wedge B_2\wedge\!\left(dC_3 - dB_2\wedge C_1\right) \nonumber \\
            &\hspace{6em}
            - \tfrac{1}{6}\,B_2\wedge B_2\wedge B_2\wedge dC_1
        \Bigr]\,.
    \label{eq:N0page}
\end{align}
The Hodge dual operator $*$ is taken with respect to the ten-dimensional metric.

\subsection{Computation of the charges}

The gauge fields of the B-R solutions are:

\begin{equation}
	\begin{split}
		C_1 &=  \left(\gamma -\frac{\mu V^2}{\hat{I}_4}\right)d\hat{t}+\left(\sqrt{1-\gamma ^2} \left(P_\theta+\alpha  V \sin ^2(\theta )\right)-\frac{\mu V^2 \left(\nu-\alpha  \mu V \sin ^2(\theta )\right)}{\sqrt{1-\gamma ^2}\hat{I}_4}\right) d\phi\\
		A_1^{3} & =(p-q)A_{-} +2 q_1 A_{+} +\frac{2 q_1 \left(p^2-q^2\right) (d\tau+d\phi P_\theta)}{\Sigma  (m-N)}+c_t^{1} dt+c_\tau^{1} d\tau+c_\phi^{1} d\phi-\frac{(dt+k)}{Z_1} \\
		A_2^{3} & =(p-q)A_{-} +2 q_2 A_{+} +\frac{2 q_2 \left(p^2-q^2\right) (d\tau+d\phi P_\theta)}{\Sigma  (m-N)}+c_t^{2} dt+c_\tau^{2} d\tau+c_\phi^{2} d\phi-\frac{(dt+k)}{Z_2} \\
		A_3^{3} & =(p+q)A_{+} +c_t^{3} dt+c_\tau^{3} d\tau+c_\phi^{3} d\phi-\frac{(dt+k)}{Z_3} \\
		A_{\pm} &= -\frac{d\tau+d\phi P_\theta}{r \mp (\alpha  \cos (\theta )+N)} \mp  \cos (\theta ) d\phi  \\
	\end{split}
\end{equation}
where $A_I^3$ is the component of the three-form $A_3$ along the volume of the torus $T^2_I$, and $c_t^I$, $c_\tau^I$ and $c_\phi^I$ are constants.
From these expressions we can deduce $C_3$ and $B_2$. Then we can compute the Page charges, by making sure we choose the right constants $c_\tau^I$ in order to make sure that $B_2$ vanishes at infinity. This constant comes from the gauge freedom of $A^3_I$. The resulting Page charges are:
\begin{center}
\begin{align}
\label{eq:BR-charges}
    N_6 &= -2N\left(\frac{2}{R_\tau}\right),\nonumber\\[6pt]
    N_4^{1} &= (2 (\gamma  N+q_1)-p+q)\left(\frac{2\,\Vt}{(2\pi)^2 l_p^3}\right),\nonumber\\[6pt]
    N_4^{2} &= (2 (\gamma  N+q_2)-p+q)\left(\frac{2\,\Vt}{(2\pi)^2 l_p^3}\right),\nonumber\\[6pt]
    N_4^{3} &= (2 \gamma  N+p+q)\left(\frac{2\,\Vt}{(2\pi)^2 l_p^3}\right),\nonumber\\[6pt]
    N_2^{1} &= -2 \left(\gamma ^2 N +\gamma  (q+q_2)+\frac{q_2 (p+q)}{N-m}\right)
            \left(\frac{2\,\Vt^2 R_\tau}{(2\pi)^4 l_p^6}\right),\nonumber\\[6pt]
    N_2^{2} &= -2 \left(\gamma ^2 N +\gamma (q+q_1)+\frac{q_1 (p+q)}{N-m}\right)
        \left(\frac{2\,\Vt^2 R_\tau}{(2\pi)^4 l_p^6}\right),\nonumber\\[6pt]
     N_2^{3} &= \Big( \lambda (N-m)-\frac{4 q_1 q_2 \left(q^2-p^2\right)}{(N-m)^3}-\frac{2 (p-q) (q+q_1+q_2)}{N-m}-\frac{4 q_1 q_2}{N-m}-2\gamma  (\gamma  N+q_1+q_2) \Big) \nonumber\\
        &\qquad \times \left(\frac{2\,\Vt^2 R_\tau}{(2\pi)^4 l_p^6}\right),\nonumber\\[6pt]
    N_0 &= \Bigg(\frac{\gamma}{(N-m)^3}  \Big((N-m)^2 \left((\lambda-2) (N-m)^2-2 p (q+2 q_1)+2 q^2\right) \nonumber\\
    &\qquad\qquad\qquad -4 q_2 \left(q_1 \left((N-m)^2-p^2+q^2\right)+p (N-m)^2\right)\Big) \nonumber\\[6pt]
    &\qquad\;  -2 \gamma ^3 N -\gamma ^2 (p+q+2 (q_1+q_2))- (p+q+2 (q_1+q_2)) \Bigg)
        \left(\frac{2\,\Vt^3 R_\tau^2}{(2\pi)^6 l_p^9}\right).
\end{align}
\end{center}
Taking the Kerr Taub-Bolt limit, which corresponds to taking $p=q$, we indeed find the Running-Kerr Taub-Bolt charges of \cite{BenaLochetPage}, with the identification \eqref{eq:correspondence KTB BR} (in this limit $\lambda=0$).

\subsection{Relations between charges}

\subsubsection{Non BPS mass relation}
In order to check whether the BPS bound is verified or not in this geometry, we would like to have a nice relation between the ADM mass and the sum of the charges. In the BPS case, the total mass is equal to the ``sum`` of each brane mass. The mass of $N_{k}$ $k$-branes is given by:
\begin{equation}
	N_{k} G_{10} M_{k} = N_{k} G_{4} V_{10}(T_{2})^{3} T_{k}V_{k}
\end{equation}
where $M_{k} \equiv T_{k}V_{k}$ is the mass of one $k$-brane, $T_{k}$ is the tension of a $k$-brane, $V_{k}$ the volume it wraps, $G_{10} =G_{4} V_{10}(T_{2})^{3} $ the link between the ten-dimensional and four-dimensional Newton constant and $V_{10}(T_{2})$ the volume of a torus with respect to the ten-dimensional metric \cite{BenaLochetPage}. Let us introduce the notation $\mathcal{Q}_{k} \equiv \dfrac{G_{4}N_{k}M_{k}}{4 \pi}$.
We obtain:
\begin{equation}
\label{eq:mass-BR-full}
    \left(4G_4 M - 2(m+N)\right)^2
    + \frac{4\left((m+N) \gamma+q-p\right)^2}{1-\gamma^2}
    = \Big(\mathcal{Q}_{6} - \sum_{I}\mathcal{Q}_{2}^{I}\Big)^2 + \Big(\mathcal{Q}_{0} + \sum_{I}\mathcal{Q}_{4}^{I}\Big)^2\,,
\end{equation}
This relation is the generalization of the one found in \cite{BenaLochetPage} for the Running-Kerr-Taub-Bolt solutions, to which it reduces when $p=q$. It holds for any values of the fluxes, including $q_1\neq q_2$.

Although the three tori are not equivalent in the bulk ($Z_3\neq Z_1,Z_2$), they are equivalent at infinity, where $Z_I\to1$: the ten-dimensional volume of each torus equals its eleven-dimensional coordinate volume,
\begin{equation}
V_{10}(T^2_I)=\frac{\hat I_4^{1/2}}{V Z_I}\bigg|_\infty V_{11}(T^2_I)=V_{11}(T^2_I)\,.
\end{equation}
Since the charges are evaluated at infinity, the inequivalence of the tori only appears in the values of the charges, not in their normalisation.

In order to investigate the BPS bound, we are interested in computing the sign of the following quantity:
\begin{equation}
\begin{split}
\label{eq:BPSboundsign}
    \Delta_{M} &\equiv (4 G_4 M)^2 - \Big(\mathcal{Q}_6 - \sum_{I}\mathcal{Q}_2^I\Big)^2 - \Big(\mathcal{Q}_0 + \sum_{I}\mathcal{Q}_4^I\Big)^2\\
    &= 4 \left( 4 G_4 M (m+N) -(m+N)^2- \frac{\left(q-p+(m+N)\gamma \right)^2}{1-\gamma^2}\right) \,,
\end{split}
\end{equation}
If the RHS is positive, then the BPS bound is respected. Expressing $\lambda$ with respect to $\gamma$ using its definition \eqref{eq:gamma},
\begin{equation}
	\lambda= -\dfrac{4 N (p-q)}{m^2-N^2} \left(\gamma +\dfrac{2 q_1^2 (p+q)}{(m-N)^3}-\dfrac{q+2 q_1}{m-N}\right)
\end{equation}
the expression \eqref{eq:BPSboundsign} greatly simplifies and reduces to:
\begin{equation}
	\Delta_{M} = \frac{4\left(\Delta_0+\Delta_\gamma\right)}{1-\gamma^2}
\end{equation}
with:
\begin{equation}
     \left\{
    \begin{array}{ll}
        \Delta_0 &= \left(1-\frac{4 q_1^2}{(m-N)^2}\right) \left(m^2-N^2-p^2+q^2\right)-\dfrac{8 q_1 (m p+N q)}{m-N},\\[1em]
        \Delta_\gamma &=4 \gamma  \Big(2 q_1 (m+N)+m p+N q\Big)-4 \gamma ^2 \left(m^2-N^2\right).
        \end{array}
\right.
\end{equation}
Without running, $\gamma=0$, the sign of the bound is the sign of $\Delta_0$. Analytically, one can check that for CTC-free solutions, the bound is satisfied. For $\gamma\neq0$, we cannot simply study \eqref{eq:BPSboundsign} as a quadratic polynomial in $\gamma$, because $\gamma$ depends on $q_1$ through \eqref{eq:lambda} and \eqref{eq:gamma}. We checked numerically that $\Delta_{M} >0$ for all the solutions of Table~\ref{tab:results N=1 selfdual}, for every value of $q_1'$ in the range free of CTCs, that is, for every $\gamma\in\,]-1,1[$. Hence the BPS bound is satisfied.

\subsubsection{BPS limit}
\label{sec:BPS limit}

The extremal limit of the base space is $r_+ = r_-$, which corresponds to $N^2 = m^2+\alpha^2+q^2-p^2$. In this limit, the periodicity relation \eqref{periodicity of phi} breaks down, but one can still ask which extremal configurations are physical. One can show analytically that the only extremal configuration that runs slower than the speed of light and is free of closed timelike curves is $m=-N$, $\alpha=0$, $p=q$, and therefore it saturates the BPS bound:
\begin{equation}
(4G_4M)^2 = \Big(\mathcal{Q}_6 - \sum_{I}\mathcal{Q}_2^I\Big)^2 + \Big(\mathcal{Q}_0 + \sum_{I}\mathcal{Q}_4^I\Big)^2 ,
\end{equation}
This is the expected result: when $p=q$ the Bobev-Ruef solution becomes a Running-Kerr-Taub-Bolt solution, and, moreover, when $m=-N$ the base space is a Taub-NUT space (written in different coordinates when $\alpha\neq0$, in which case it is not extremal since $r_+-r_-=2\alpha$), and the solution is a supersymmetric fluxed Taub-NUT solution. The study of this limit is thus the one done for the Running-Kerr-Taub-Bolt solutions in Section~5 of \cite{bena2025positivespininducedquadrupolemoment}, and its charges and BPS mass relation are discussed in \cite{BenaLochetPage}.

\subsection{Quartic invariant and entropy}

To discuss the entropy of a black hole with the same charges, the relevant quantity to compute is the $E_{7(7)}$-quartic invariant \cite{Bertolini_2000,Bena:2004tk}
\begin{equation}
\label{eq:E77}
\begin{split}
    E_{7(7)} & \equiv -\left(x_{12}y_{12} + x_{34}y_{34} + x_{56}y_{56} + x_{78}y_{78}\right)^2- 4\bigl(x_{12}x_{34}x_{56}x_{78} + y_{12}y_{34}y_{56}y_{78}\bigr)\\
    &\quad + 4\bigl(
        x_{12}x_{34}y_{12}y_{34} + x_{12}x_{56}y_{12}y_{56}
        + x_{12}x_{78}y_{12}y_{78} + x_{34}x_{56}y_{34}y_{56}\\
    &\qquad  \quad + x_{34}x_{78}y_{34}y_{78} + x_{56}x_{78}y_{56}y_{78}
    \bigr)\,.
\end{split}
\end{equation}

For the D0–D2–D4–D6 system the values of the parameters $x_{ij}, y_{ij}$ are according to \cite{Bertolini_2000,Bena:2004tk} (see also \cite{BenaLochetPage} for the signs):
\begin{equation}
\label{eq:E77-charges}
\begin{aligned}
    &x_{12} = {N}_2^1\,,\quad x_{34} = {N}_2^2\,,\quad x_{56} = {N}_2^3\,,\quad x_{78} = {N}_6\,,\\
    &y_{12} = -{N}_4^1\,,\quad y_{34} = -{N}_4^2\,,\quad y_{56} = -{N}_4^3\,,\quad
    y_{78} = {N}_0\,.
\end{aligned}
\end{equation}
Using the charges \eqref{eq:BR-charges}, one finds, exactly as for the Running-Kerr-Taub-Bolt solutions \cite{BenaLochetPage},
\begin{equation}
\label{eq:E77-J}
	E_{7(7)} = -16\,(G_4J)^2
\end{equation}
for any values of the fluxes, including $q_1\neq q_2$ (with $G_4J$ given in the footnote of Section~\ref{sec:multipoles}). In particular, $E_{7(7)}$ vanishes exactly when the angular momentum vanishes, as in the Taub-NUT limit discussed in Section~\ref{sec:BPS limit}.

In the conventions of Chow and Comp\`ere \cite{Chow:2014cca}, $E_{7(7)}$ is proportional to the Cayley hyperdeterminant of the charges, $E_{7(7)}=16\,\Delta_{CC}$ \cite{BenaLochetPage}, and \eqref{eq:E77-J} becomes $\Delta_{CC}=-(G_4J)^2$. Hence, as for the Running-Kerr-Taub-Bolt solutions, a black hole with the same charges and angular momentum would have a vanishing inner-horizon entropy, and its outer-horizon entropy, $S_+=4\pi\sqrt{F-(G_4J)^2}$, where $F$ is the $F$-invariant of \cite{Chow:2014cca}, is real only if $F\geq (G_4J)^2$. Whether the Bobev-Ruef solitons have the charges of a black hole with a regular horizon is therefore not determined by \eqref{eq:E77-J} alone.\\
\vspace{1em}

\noindent\textbf{Is there a corresponding black hole?}
\label{sec:BH bound}\\
To answer this question, we follow the procedure of \cite{BenaLochetPage}. We reduce the solution to four dimensions in the conventions of \cite{Chow:2014cca} and read off its mass and its electric and magnetic charges, $(Q_I,P^I)$. The mass and the angular momentum obtained in this way agree with \eqref{M self dual} and \eqref{J self dual}. We then invert the charge map of the general black hole of \cite{Chow:2014cca} at fixed $(Q_I,P^I)$, with equal charging parameters for the first two tori, since these tori are equivalent when $q_1=q_2$. This gives a one-parameter family of black holes with the same charges, whose mass decreases monotonically to the extremal value $M_{\rm ext}(Q,P)$ (we found no other branch). A black hole with the charges of the soliton can therefore exist only if $M\geq M_{\rm ext}(Q,P)$, and, because $\Delta_{CC}=-(G_4J)^2$, this condition does not depend on the angular momentum \cite{BenaLochetPage}.

The results for the non-running solutions of Table~\ref{tab:results N=1 selfdual} are given in Table~\ref{tab:BH bound}: the mass of the soliton always lies a few percent below the extremal black-hole mass with the same charges. This remains true for running solutions: along the ranges of fluxes free of CTCs of the first and last lines of Table~\ref{tab:results N=1 selfdual}, with running speeds up to $|\gamma|\simeq0.97$, we find $0.956\leq M/M_{\rm ext}\leq0.993$. Hence, as for the Running-Kerr-Taub-Bolt solutions \cite{BenaLochetPage}, there is no black hole, static or rotating, with the mass and the charges of these solitons.

\begin{table}[h!]
    \centering
    \begin{tabular}{|c|c|c|c|c|c|}
        \hline
        $\ell$ & $p'$ & $q_1'$ ($\gamma=0$) & $G_4M$ & $G_4M_{\rm ext}$ & $M/M_{\rm ext}$ \\
        \hline
        1 & $2.5$ & $-0.5998$ & 4.2046 & 4.3659 & 0.963 \\
        \hline
        2 & $2.5$ & $-1.2459$ & 9.4033 & 9.6395 & 0.976 \\
        \hline
        3 & $2.3$ & $-1.3644$ & 8.2352 & 8.2984 & 0.992 \\
        \hline
        1 & $-2.5$ & $0.6076$ & 3.8650 & 3.9982 & 0.967 \\
        \hline
    \end{tabular}
    \caption{Mass of the non-running solitons of Table~\ref{tab:results N=1 selfdual} ($N=1$, $m'=1.8$, $q_1=q_2$) and extremal mass of the black holes with the same charges, in units $G_4=N=1$. In all cases $M<M_{\rm ext}$: no black hole has the charges and the mass of these solitons.}
    \label{tab:BH bound}
\end{table}

\section{Conclusions}
\label{sec:conclusion}

In this paper we calculated the spin-induced quadrupole moment and the brane charges of the Bobev-Ruef solitons \cite{Bobev:2009kn}. These smooth horizonless solutions of eleven-dimensional supergravity are built by adding fluxes on a Euclidean Einstein-Maxwell base space, which generalizes the Euclidean-Kerr-Taub-Bolt space by an electric and a magnetic parameter. We identified the base spaces that give solutions with flat four-dimensional asymptotics. Because of the extra parameter $\delta=q^2-p^2$, these base spaces form lines in parameter space, labeled by an integer $\ell$, instead of the isolated points found for the Kerr-Taub-Bolt base spaces.

We computed the mass, the angular momentum and the quadrupole moment of these solutions, and found that the quadrupole moment takes the simple form $M_2=M(\alpha^2-b^2)$, where $b$ is the displacement of the center of mass. The same relation holds for the Running-Kerr-Taub-Bolt solutions \cite{bena2025positivespininducedquadrupolemoment}. For all the solutions free of closed timelike curves that we studied, the quadrupole moment is positive in a finite range of fluxes, which contains the solution that does not run, and it becomes negative and diverges when the running speed approaches the speed of light. Together with the rotating topological stars of \cite{HPS}, which can also have a positive quadrupole moment in a small region of their parameter space, our results show that a positive spin-induced quadrupole moment is not a peculiarity of the Running-Kerr-Taub-Bolt solutions \cite{bena2025positivespininducedquadrupolemoment}, but appears in several families of rotating smooth horizonless solutions. 

The Bobev-Ruef solitons also differ in two interesting ways from the Running-Kerr-Taub-Bolt solutions. First, their multipole ratio, $R=M_2M/J^2$, can be much larger, reaching $R\simeq2.07$ in one of our examples. Second, some of their base spaces have no region of $(-,-,-,-)$ signature. Thus, the positive quadrupole moment does not require an ambipolar base space.

We also computed the Page charges of these solutions. We found a mass formula that generalizes the one of the Running-Kerr-Taub-Bolt solutions, and showed that the BPS bound is respected. The only extremal configuration that is free of closed timelike curves and runs slower than the speed of light is the supersymmetric fluxed Taub-NUT solution, which saturates this bound. We also found that the quartic invariant of the charges is related to the angular momentum by $E_{7(7)}=-16(G_4J)^2$ for all values of the fluxes. Using these charges, we showed that the mass of the solitons we studied lies a few percent below the mass of the extremal black hole with the same charges: no black hole has the charges and the mass of these solitons.

There are several open questions that our results raise.
First, we restricted our analysis to $q_1=q_2$, $\ell=1,2,3$, $N>0$ and $m>0$. The mass and the angular momentum for $q_1\neq q_2$ are given in Section~\ref{sec:multipoles}, and it would be interesting to explore the full space of solutions, as well as the large-$\ell$ limit, in which the multipole ratio of the Running-Kerr-Taub-Bolt solutions tends to zero \cite{bena2025positivespininducedquadrupolemoment}.

Second, the relation $M_2=M(\alpha^2-b^2)$ holds for both families of solutions we studied, and it seems to be the first sign of a much simpler structure. Indeed, we also checked that, for both families and for running as well as non-running solutions, the octupole moment satisfies $M_3=2b\,M_2$ (for the non-running Running-Kerr-Taub-Bolt solutions, this can also be seen from the expression of $M_3$ given in \cite{bena2025positivespininducedquadrupolemoment}). These two relations are exactly those satisfied by the mass multipoles of two point masses, $\frac{M}{2}\big(1\mp\frac{b}{\alpha}\big)$, located on the axis at the foci, $z=\pm\alpha$, of the prolate spheroidal coordinates of Section~\ref{sec:multipoles}, measured from their center of mass. This is reminiscent of the Kerr black hole, whose multipole moments, $M_l+iS_l=M(ia)^l$ \cite{Hansen:1974zz}, are formally those of a source located at the imaginary positions $z=\pm ia$, the Euclidean rotation parameter $\alpha$ playing here the role of $ia$. It would be interesting to understand whether all the mass and current multipoles of these solutions admit such a simple description, and whether this also holds for other rotating solutions built from Euclidean rotating spaces, such as the rotating topological stars of \cite{HPS}, whose quadrupole moment also takes a simple form. More generally, it would be interesting to see whether the complicated-looking multipole moments of other solutions simplify when they are written in terms of the displacement of their center of mass: for example those of the most general non-extremal black hole of the STU model \cite{Bena:2020uup}, of the Kerr-NUT solution \cite{Mukherjee:2020how}, of black holes in higher-derivative gravity \cite{Cano:2022wwo}, or of supersymmetric and almost-BPS microstate geometries \cite{Bena:2020see,Bianchi:2020bxa,Bianchi:2020miz,Ganchev:2022vrv,Bah:2021jno} (see also \cite{Mayerson:2022ekj} for the multipole moments of general stationary spacetimes).

Third, as for the Running-Kerr-Taub-Bolt solutions \cite{BenaLochetPage}, the single-bolt solutions we studied lie below the black-hole bound, and the relation $E_{7(7)}=-16(G_4J)^2$ implies that the extremal black hole with their charges and angular momentum would have zero entropy. This supports the idea, suggested in \cite{bena2025positivespininducedquadrupolemoment}, that spinning single-bolt solutions are obstructed from reaching the black-hole regime, and it would be important to prove it, for instance by deriving the identity $E_{7(7)}=-16(G_4J)^2$ directly from the properties of the bolt. There are two natural ways to go beyond this obstruction. The first is to use solution-generating techniques: the rotating topological stars of \cite{HPS} were obtained by acting with dualities on the ``Euclidean-Kerr-Taub-Bolt times time'' solution, and the same dualities could be applied to the Bobev-Ruef solitons to produce solutions in new duality orbits, which might have black-hole charges. 
The second is to construct spinning solutions with several bolts, or with unequal fluxes, and with the charges of non-extremal black holes, as was done for static solutions \cite{Bah:2022pdn,Heidmann:2021cms,Bah:2021owp,Bah:2021rki,Heidmann:2022zyd,Bah:2022yji,Bah:2023ows,Heidmann:2023kry}, and to see whether the unusual spin-induced quadrupole moment we found is a generic feature of such solutions.

Fourth, it would be interesting to determine whether positive spin-induced quadrupole moments are ubiquitous in other rotating smooth solutions with four-dimensional asymptotics, such as the bubbling bolts of Bossard and Katmadas \cite{Bossard:2014yta,Bossard:2014ola,Bena:2015drs,Bena:2016dbw,Bossard:2017vii}, and whether, as the Bobev-Ruef solitons suggest, they can arise without an ambipolar base space.

\noindent{\bf Acknowledgments:} \\
I would like to thank Iosif Bena, Pierre Heidmann and Guillaume Bossard for interesting discussions. I am particularly grateful to Iosif for suggesting this project, for insightful discussions and for his comments on the draft.
This work was supported by ENS Paris-Saclay doctoral grant.
\noindent{\bf AI statement:}\\
In the preparation of this work, I used the large language model Claude Opus 5.5 (Anthropic). It was used to check the analytic and numerical computations independently, to search the literature, to write the Python scripts producing the figures, and to rephrase parts of the text. I have verified all the results, derivations and text, and I take full responsibility for the content of this paper.

\appendix

\section{Supergravity background}
\label{Appendix Supergravity}

This appendix summarises our supergravity conventions and the brane configuration underlying the solutions of Section~\ref{sec:fluxes}. They are the same as in~\cite{bena2025positivespininducedquadrupolemoment}, and we recall them here to keep the paper self-contained.

Our starting point is eleven-dimensional supergravity, with bosonic action~\cite{Chakraborty:2025ger}
\begin{equation}
\label{action sugra}
	S_{11} = \frac{1}{16 \pi G_{11}} \int  R *_{11} 1 -\frac{1}{2} F_4 \wedge *_{11} F_4 - \frac{1}{6} F_4 \wedge F_4 \wedge A_3 \, ,
\end{equation}
where $F_4 = dA_3$, and the subscript on $*_{11}$ recalls that the Hodge dual is built with the eleven-dimensional metric. The last term is the Chern--Simons coupling. Varying~\eqref{action sugra} with respect to $A_3$ gives the equation of motion
\begin{equation}
	d  *_{11} F_4 = -\frac{1}{2} F_4 \wedge F_4  \, .
\end{equation}
We split the eleven-dimensional spacetime as $\mathcal{M}_5\times T^2_1\times T^2_2\times T^2_3$. The five-dimensional part $\mathcal{M}_5$ consists of the time direction and a four-dimensional Euclidean base, which contains the circle $S^1$, parametrised by $\tau$, along which one reduces from eleven to ten dimensions. Here the base is the Bobev-Ruef base~\eqref{eq:BRbase}.

The branes are listed in Table~\ref{tab:brane setup}. There are three M5 branes, each wrapping $S^1$ and two of the tori, and three M2 branes, each wrapping a single torus. Every brane is smeared along the compact directions it does not wrap. This configuration of branes is BPS, but the solutions considered here are not supersymmetric, because the base is Ricci-flat but not hyper-K\"ahler.

\begin{table}[h!]
    \centering
    \begin{tabular}{|c|c|c|c|c|c|c|}
        \hline
         & $0$ & $1$ & $2\hspace{1mm}3\hspace{1mm}4$ & $5\hspace{1mm}6$ & $7\hspace{1mm}8$ & $9 \hspace{1mm} 10$ \\
        \hline
         & $\mathbb{R}_{t}$ & $S^1_{\tau}$ & $\mathcal{M}_3$ & $T^2_1$ & $T^2_2$ & $T^2_3$ \\
        \hline
        $M5_1$ & $\times$ & $\times$ & $\perp$ & -- & $\times$ & $\times$ \\
        \hline
        $M5_2$ & $\times$ & $\times$ & $\perp$ & $\times$ & -- & $\times$ \\
        \hline
        $M5_3$ & $\times$ & $\times$ & $\perp$ & $\times$ & $\times$ & -- \\
        \hline
        $M2_1$ & $\times$ & -- & $\perp$ & $\times$ & -- & -- \\
        \hline
        $M2_2$ & $\times$ & -- & $\perp$ & -- & $\times$ & -- \\
        \hline
        $M2_3$ & $\times$ & -- & $\perp$ & -- & -- & $\times$ \\
        \hline
    \end{tabular}
    \caption{Directions spanned by each brane: $\times$ marks a direction along the worldvolume, $\perp$ a transverse direction, and -- a direction along which the brane is smeared. The branes are labelled so that M2$_I$ and M5$_I$ are the electric and magnetic sources of the same gauge field $A^I$.}
    \label{tab:brane setup}
\end{table}

In the floating-brane ansatz~\cite{Bena:2009fi}, the three-form potential takes the form
\begin{equation}
\label{eq:A3 ansatz}
	A_3 = \sum_{I=1}^{3} A^{I} \wedge dz^{2I+3} \wedge dz^{2I+4} \,, \qquad
	A^{I} = - \varepsilon\, Z^{-1}_I \left(dt+k\right) + B^{I} \, ,
\end{equation}
where $(z^{2I+3},z^{2I+4})$ are the coordinates on $T^2_I$, i.e.\ directions $5,\dots,10$ of Table~\ref{tab:brane setup}, and $Z_I$ and $k$ are the warp factors and the rotation one-form of the metric~\eqref{eq:11Dmetric}. The one-forms $B^I$ live on the base, and their field strengths $\Theta^I \equiv dB^I$ are the magnetic fluxes threading the bolt. The sign $\varepsilon = \pm 1$ selects self-dual or anti-self-dual fluxes, $\Theta^I = \varepsilon *_4 \Theta^I$, where $*_4$ is the Hodge dual of the base. It is called the floating-brane ansatz because M2-brane probes parallel to the M2$_I$ feel no force, and the equations of motion reduce to a linear system on a Ricci-flat base.

Compactifying on the three tori gives five-dimensional $\mathcal{N}=2$ supergravity coupled to two vector multiplets, known as the STU model or $U(1)^3$ supergravity~\cite{Chakraborty:2025ger}. Here this step is simple: each brane either wraps a torus or is smeared along it, so no field depends on the torus coordinates. The five-dimensional action is
\begin{equation}
	S_5 = \frac{1}{16 \pi G_5} \int R_5 *_5 1 - \frac{1}{2}Q_{IJ}\, d X^{I} \wedge *_5 d X^{J}-  \frac{1}{2} Q_{IJ}\, F^{I} \wedge *_5 F^{J}- \frac{C_{IJK}}{6}\, F^{I} \wedge F^{J} \wedge A^K \, ,
\end{equation}
where $F^I\equiv dA^I$, $C_{IJK} = |\epsilon_{IJK}|$, $G_5 = G_{11}/V_{T^6}$ with $V_{T^6}$ the volume of the three tori, and $Q_{IJ}=\mathrm{diag}\big((X^1)^{-2}, (X^2)^{-2},(X^3)^{-2}\big)$. The scalars obey $X^1 X^2 X^3 = 1$, so only two of them are independent.
They measure the sizes of the tori; comparing with~\eqref{eq:11Dmetric},
\begin{equation}
	X^I = \frac{Z}{Z_I}\,, \qquad Z \equiv \left(Z_1 Z_2 Z_3\right)^{1/3} .
\end{equation}
Because $X^1X^2X^3=1$, the total volume of $T^6$ is constant and the five-dimensional metric is just the first two terms of~\eqref{eq:11Dmetric},
\begin{equation}
	ds_5^2 = -Z^{-2}\left(dt+k\right)^2 + Z\, ds^2_{\mathrm{EKTB}} \,.
\end{equation}

Setting $X^1=X^2=X^3=1$ and $A^1=A^2=A^3\equiv A$ is a consistent truncation. For the solutions of this paper it corresponds to $Z_1=Z_2=Z_3$, i.e.\ to equal fluxes $2q_1=2q_2=q+p$. The action becomes
\begin{equation}
	S_5^{\mathrm{min}} = \frac{1}{16 \pi G_5} \int  R_5 *_5 1 -  \frac{3}{2} F\wedge *_5 F- F \wedge F \wedge A  \, ,
\end{equation}
with $F\equiv dA$. This is the bosonic sector of minimal five-dimensional supergravity: Einstein--Maxwell theory with a Chern--Simons term whose coefficient is fixed by supersymmetry. A different normalisation is also common in the literature,
\begin{equation}
	S_5^{\mathrm{min}} = \frac{1}{16 \pi G_5} \int R_5 *_5 1 -  \frac{1}{2} F\wedge *_5 F- \frac{1}{3 \sqrt{3}}F \wedge F \wedge A  \, .
\end{equation}
The two are related by the rescaling $A_{\mathrm{here}}= A_{\mathrm{lit}}/\sqrt{3}$, which puts the Maxwell term in canonical form. All the numerical results of this paper are obtained in minimal supergravity, which reduces the number of free parameters.

\begin{adjustwidth}{-1mm}{-1mm}
\bibliographystyle{utphys}
\bibliography{biblio_BR}
\end{adjustwidth}

\end{document}